\documentclass{article}
\usepackage{frascatiphys,here,graphicx,subfigure}
\begin{document}
\title{$\pi^o$ ELECTROPRODUCTION AND TRANSVERSITY}
\author{
Simonetta Liuti, Saeed Ahmad        \\
{\em University of Virginia, Charlottesville, VA 22904, USA} \\
Gary~R.~Goldstein        \\
{\em Tufts University, Medford, MA 02155, USA} \\
Leonard Gamberg        \\
{\em Pennsylvania State University, Penn State Berks, Reading, PA 19610, USA}
}
\maketitle

\baselineskip=11.6pt
\begin{abstract}
Exclusive $\pi^o$ electroproduction and related processes are suggested to 
investigate the chiral odd transversity distributions of quarks in the transversely 
polarized nucleon, $h_1(x)$,  and its first moment, the tensor charge.  
The connection between a description based on partonic degrees of freedom, 
given in terms of generalized parton distributions, and Regge phenomenology is explored.
\end{abstract}
\baselineskip=14pt
\section{Introduction}
Most of the information on the partonic structure of hadrons 
has been traditionally obtained through inclusive deep inelastic experiments. 
With an appropriate selection of 
probes and reactions, 
accurate measurements conducted through the years  
allowed one to map out in detail the different components of proton 
structure, the Parton Distribution Functions (PDFs) 
in a wide kinematical region of  
the four-momentum transfer, $Q^2$, and of the longitudinal momentum 
fraction of the proton's momentum, $x_{Bj}$.
An inclusive/exclusive connection in lepton-nucleon
scattering was advocated \cite{BroLep}, although 
studies of the partonic structure through exclusive measurements remained 
ambiguous in establishing the regime of 
four-momentum transfer in which such a description should be valid.
A new avenue was recently suggested 
in view of the factorization properties of new types 
of exclusive processes, 
namely Deeply Virtual Compton Scattering (DVCS), Deeply Virtual Meson  
Production (DVMP), and related crossed-channel reactions. 
The soft matrix elements, identified 
with Generalized Parton Distributions (GPDs), 
can describe both 
the intrinsic motion of partons and their coordinate 
space representation (see \cite{Die_rev,BelRad} for
reviews). 
It is at present an outstanding problem to be able to reconcile 
and properly connect the newly suggested QCD-based picture with the hadronic 
description of hard exclusive reactions based on Regge theory. The latter
is well known to predict a vast number of observations at large $s$, and small $t$,
corresponding to the $x_{Bj} \rightarrow 0$ limit of inclusive Deep Inelastic Scattering
(DIS). Initial studies of the correspondence between the regime
expected to be dominated by Regge exchanges and the partonic description in DVCS and DVMP 
were performed in \cite{LonSzc1,LonSzc2} where it was claimed that 
the ``leading Regge trajectories should dominate the amplitudes for exclusive leptoproduction''.
Therefore one might think of the hard exclusive process as probing the deep inelastic structure
of a $t$-channel exchange target, or the mesonic cloud. 
On the other side, it was pointed out in Ref.\cite{MueKum} 
that the model used in \cite{LonSzc1,LonSzc2} might disagree with the perturbative QCD 
scaling violation pattern, which was recently observed in a particular (large $W^2$) 
kinematical regime. 
A formalism using the conformal moments of the nucleon GPDs was instead 
introduced 
with a non-perturbative input based on Regge ans\"{a}tze.

Whether or not specific models are seemingly able to reproduce the current trend of data 
it is important to determine the physical origin of the hadronic structure that is detectable
with hard exclusive scattering experiments aiming to shed light on the complementary picture
where hadron structure at this transition regime emerges through QCD 
dynamics.

In this context we analyze a specific class of exclusive reactions, namely $\pi^o$ 
electroproduction with the goal of obtaining a relation between experimental 
observables and the tensor charge. 

\section{Tensor Charge and Transversity}
The chiral odd transversity distribution of quarks in the transversely 
polarized nucleon, $h_1(x)$, and its first moment, the tensor charge, $\delta q$,  
are defined as
\begin{eqnarray}
\langle P S \mid \overline{\psi} \, \sigma^{\mu\nu} \, \psi \mid PS \rangle 
& = & \delta q \; \overline{U}(P,S) \sigma^{\mu\nu} U(P,S)
\nonumber \\ 
& \equiv &   \delta q \; %
2 \left( P^\mu S^\nu - P^\nu S^\mu \right),  
\end{eqnarray}
and 
\begin{eqnarray}
h_1(x_{Bj},Q^2) = \Phi^\Gamma = \int dp^- \, d^2{\bf p_T} \, 
\rm{Tr} \left\{ \Gamma \Phi \right\} \mid_{x_{Bj}P^+=p^+},
\end{eqnarray}
with $\Gamma = i \sigma^{i \, +} \gamma_5$, and 
$\Phi(x; P,S)$ being the correlation function defining the matrix element for the 
DIS process \cite{JafJi}.
%

\noindent 
Notice that in the helicity basis \cite{BogMul} $h_1$ corresponds to the off-diagonal
quark chirality matrix elements
\begin{equation}
h_1 = \Phi_{+L,-R} + \Phi_{-R,+L} = q^\uparrow(x_{Bj},Q^2) - q^\downarrow(x_{Bj},Q^2)
\label{h1_helic}
\end{equation}
originally introduced in \cite{GolMor,RalSop}
where the subscripts $\Lambda \lambda, \Lambda^\prime \lambda^\prime$ refer to 
the helicities of the proton ($\Lambda$, $\Lambda^\prime$), and of the struck quark 
($\lambda$, $\lambda^\prime$). The connection between tensor charge and transversity 
is given by \cite{JafJi}:
\begin{equation}
\delta q = \int\limits_0^1 h_1(x_{Bj},Q^2) \, dx_{Bj}
\end{equation}
In a non-relativistic scenario $h_1(x_{Bj},Q^2)$ would coincide 
with the distribution of longitudinally polarized quarks in a longitudinally
polarized proton, $g_1(x_{Bj},Q^2)$.  
Its dynamical origin is therefore related to the relativistic motion of 
quarks in the nucleon.

\begin{figure}[H]
\begin{center}
~\subfigure{\includegraphics[scale=0.3]{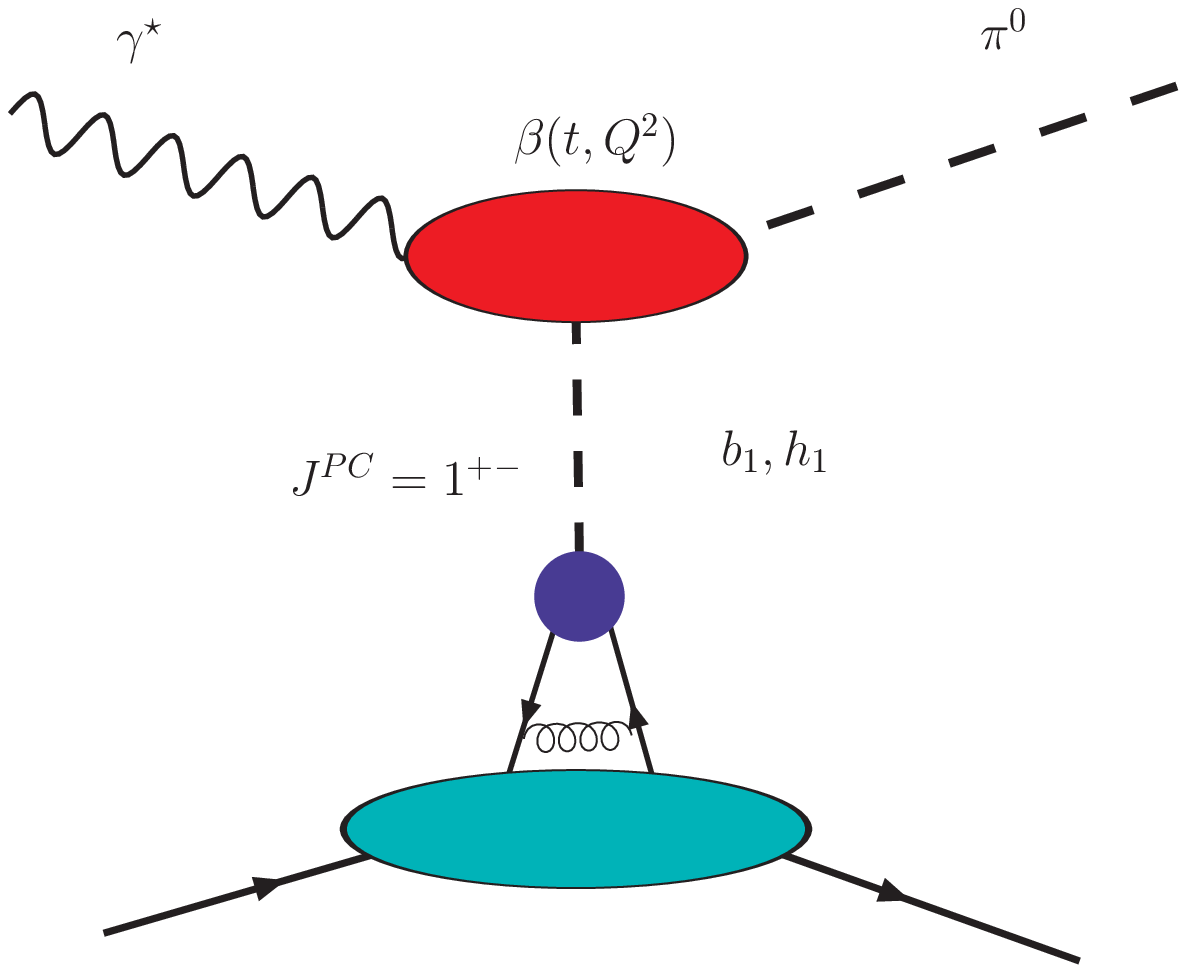}}\hspace{1cm}
~\subfigure{\includegraphics[scale=0.3]{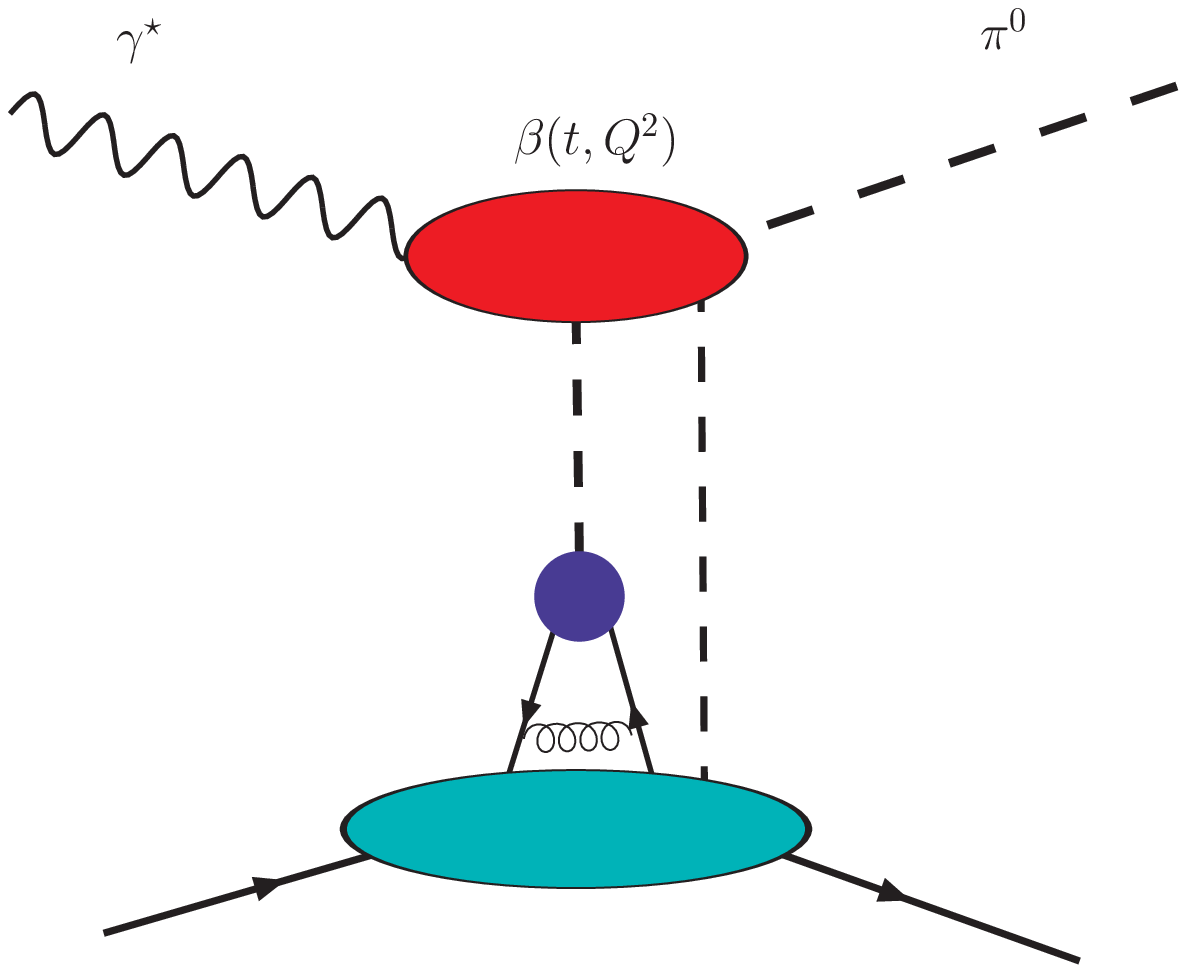}}
\end{center}
\caption{{\it The factorized
 Regge pole contribution to $\pi^0$ scattering is
depicted.}}
\label{regge}
\end{figure}

\begin{figure}[H]
\begin{center}
~\subfigure{\includegraphics[scale=0.3]{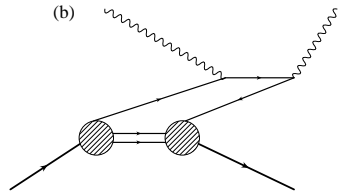}}\hspace{1cm}
\end{center}
\caption{{\it Leading order diagram for DVCS in the ERBL \cite{BroLep,ERBL} region 
where a $q \overline{q}$ pair is first produced from the nucleon and subsequently 
interacts with the photons.}}
\vspace{-0.4cm}
\label{topt}
\end{figure}

Many attempts have been made to connect the tensor charge 
to measurable processes, the most 
successful of which have been through semi-inclusive DIS \cite{Pap}. 
Various theoretical 
approaches to modeling these quantities have been taken: from QCD Sum rules \cite{HeJi} 
to Lattice QCD \cite{Goc}
to phenomenological studies  \cite{Ans}.
One particular approach to predicting the nucleon's tensor 
charges, $\delta u$ and $\delta d$ has been the work of
Gamberg and Goldstein~\cite{Gamberg:2001qc}. The tensor charges were
calculated using approximate ${\rm SU(6)\otimes O(3)}$
symmetry among the light axial vector mesons, an
 axial vector dominance hypothesis,  and a
generalization with re-scattering. 
%
This formalism
which is based upon $t$-channel exchange is reminiscent of
Regge Cut models on the one hand, depicted in Fig.~\ref{regge}, and the
the large $s$, small $t$ (large $u$) 
$q\overline{q}$ pair (or to a t-channel exchange) 
or ERBL region of DVMP on 
the other hand, depicted in Fig.~\ref{topt}. 

This scheme yielded values for $u$ and $d$ quark tensor charges dependent on
transverse momentum factors interpreted as
transverse momentum transfer $\Delta_\perp^2$ of the quarks in the nucleon.
$\Delta_\perp^2$ dependence suggests that re-scattering loop
corrections must be present in single spin asymmetries in exclusive
pion production~\cite{Goldstein:1973xn,Gamberg:2001qc}. 
This approach left two open questions. First, the 
poles in momentum transfer $t$ are quite far removed from the relevant 
$t\rightarrow 0$ limit. Secondly, the charges depend on average values 
of the constituent transverse momenta, $\langle k_T^2\rangle$. 
To address these questions and to develop a deeper understanding of the 
partonic underpinnings of transversity, we addressed them 
using the connection between GPDs and transversity recently studied in 
\cite{DieHag}. 

Note that
$H_T(x,\xi,t)$,  
the Generalized Parton Distribution for transversity can be written as
the off forward quark-target scattering amplitude, $A_{\Lambda \lambda,\Lambda^\prime\lambda^\prime}$,
\begin{equation}
H_T(X,\zeta,t) = A_{++,--} + A_{--,++},
\label{HT_helicity}
\end{equation} 
where $H_T(x,\xi,t)$ 
 has the following properties:
\begin{eqnarray}
\int H_T(x,\xi,t) \, dx = A_{T,10}(t) \\
H_T(x,0,0) = h_1(x)  
\end{eqnarray}
and the form factor $A_{T,10}(t)$ gives the tensor 
charge 
for $t \rightarrow 0$.
\footnote{Notice that $A_{\Lambda \lambda,\Lambda^\prime\lambda^\prime}$, reduces 
to $\Phi_{\Lambda \lambda,\Lambda^\prime\lambda^\prime}$, Eq.(\ref{h1_helic}), 
in the forward limit.}

In order to investigate more extensively the tensor charge and 
the possible mechanisms at work in both the partonic and hadronic pictures
exclusive electroproduction of  
$\pi^0$ and $\eta$ on both proton and neutron targets, 
can be used where a more direct connection between theoretical quantities and 
observables can be established. 

\section{Extraction of Tensor Charge from Data}
We propose a dynamical  mechanism for the process 
$\gamma^* P \rightarrow \pi^0 (\eta) P^\prime$ 
that allows for a direct extraction of the tensor charge from 
experiment. 
The cross section for $\pi^o$ electroproduction reads
\begin{eqnarray}
\frac{d \sigma}{dt \, d \phi} & \propto & L_{\mu \nu}^{h=\pi^o} W^{\mu \nu} 
\nonumber \\%
& = &  \left(\frac{d \sigma_T}{dt} + \epsilon \frac{d \sigma_L}{dt} \right) +%
\epsilon \frac{d \sigma_{TT}}{dt} \cos 2 \phi + \sqrt{2 \epsilon (\epsilon +1)}%
\frac{d \sigma_{LT}}{dt} \cos \phi.   
\end{eqnarray}
$L_{\mu \nu}^{h=\pi^o}$ is the leptonic tensor, or polarization 
density matrix, and
\begin{equation}
W^{\mu \nu} =  \sum_f J_\mu J^*_\nu \delta(E_i-E_f)
\end{equation}
where the sum is carried out over all final states, and $J_\mu$ is the matrix element
of the proton electromagnetic current operator between the initial and final states.
Note that the quantity
\begin{eqnarray}
\frac{d \sigma_{TT}}{dt} & = & 
W_{yy} - W_{xx} \equiv 2 \Re e (J_1 J_{-1}* )
\end{eqnarray}
can also be written in terms of the helicity amplitudes introduced in Section 2.
$d \sigma_{TT}/dt$ enables us to access 
the tensor charge by taking the $t \rightarrow 0$ limit of the 
only non-flip helicity amplitude for the process. 
This is in fact proportional
to a combination of unnatural parity exchanges (see Fig.\ref{regge}) that
provide the quantum numbers
in the $t$-channel that are necessary to produce a chirality flip and emit the 
$\pi^o$.   
Notice that other quantities that provide similar information 
such as the target transverse polarization asymmetry, 
$A_{UT}$, can be
considered. These will be discussed in detail in a forthcoming publication.

In this contribution we present preliminary results using both the hadronic and the 
partonic descriptions, the latter obtained by 
implementing a recent GPD parametrization \cite{AHLT1}.
A practical extraction of the tensor charge can be obtained  
by noticing that in either the hadronic or in the partonic, 
GPD-type, description of {\it e.g.} 
$d \sigma_{TT}/dt$, 
the tensor charges for the different isospin components might be treated
as parameters related to the normalization of $H_T$. Therefore
our models can be used to constrain the range of values allowed by the data.
\begin{figure}[H]
\begin{center}
~\subfigure{\includegraphics[scale=0.26]{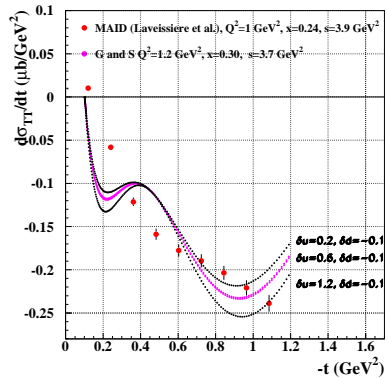}}
\end{center}
\vspace{-0.4cm}
\caption{\it{Experimental determination of the tensor charge for $u$ and $d$ quarks, 
$\delta u$ and $\delta d$, using
MAID data on $d\sigma_{TT}/dt$ ~\cite{maid}.
The suggested extraction method is explained in the text.}
\vspace{-0.4cm}
\label{maid}}
\end{figure}
As an example in Fig.~\ref{maid} we show a comparison 
of our Regge model with the MAID~\cite{maid}  
parametrization 
of pion electroproduction data at $s=$ 4 GeV$^2$ and $Q^2 \approx $ 1
GeV$^2$. 
The different curves were obtained by varying 
the values of the tensor charge entering 
the normalization of the different $t$-channel exchanges.
We expect a variety of new flavor sensitive 
observables to be extracted from data  in the near 
future using both unpolarized data and asymmetries from transversely polarized
proton and deuteron data on $\pi^o$ and $\eta$ production at the 
higher values of $s$ available at Jefferson Lab.
This analysis promises to be a 
rich area of experimental exploration in the near future.

\section{Conclusions}
We presented our preliminary results on 
an alternative method to extract the tensor charge, and its related GPDs 
from experimental observables such as the structure 
function  $d \sigma_{TT}/dt$ in unpolarized exclusive 
$\pi^o$ electroproduction. 
Our study uses 
both a Regge model, and a diquark scheme for the transverse GPD, 
$H_T$, which include as parameters the isospin dependent tensor charges.  

A feasibility study with members of
the Hall B collaboration at Jefferson Lab is in progress.  

%

%

%
\end{document}